\documentclass[pra,showpacs,twocolumn,amsmath,amssymb,floatfix]{revtex4}
\usepackage{graphicx,bm,color}
\begin{document}
\title{Analog control of open quantum systems under arbitrary decoherence}
\date{\today}
\author{Jens Clausen}
\email{Jens.Clausen@weizmann.ac.il}
\author{Guy Bensky, and Gershon Kurizki}
\affiliation{Department of Chemical Physics,\\
             Weizmann Institute of Science,\\
             Rehovot,
             76100, Israel}
\begin{abstract}
We derive and investigate a general non-Markovian equation for the
time-dependence of a Hamiltonian that maximizes the fidelity of a desired
quantum gate on any finite-dimensional quantum system in the presence of
arbitrary bath and noise sources.
The method is illustrated for a single-qubit gate implemented on a three-level
system.
\end{abstract}
\maketitle
\section{Introduction}
The quest for strategies for combatting decoherence is of paramount
importance to the control of open quantum systems, particularly for quantum
information operations \cite{krausNielsen}.
A prevailing \emph{unitary} strategy aimed at suppressing decoherence is 
dynamical decoupling (DD) \cite{vio98,KLi05,uhr07}, which
consists, in the case of a qubit,
in the application of strong and fast pulses
alternating along orthogonal Bloch-sphere axes, e.g.,  $X$ and $Z$.
In the frequency domain,  where the decoherence rate can be described as overlap
between the spectra of the pulse-driven (modulated) system and the bath
\cite{kof01}, DD is tantamount to shifting the driven-system
resonances beyond the bath cutoff frequencies. The DD efficacy can be enhanced
for \emph{certain} bath spectra upon choosing the timings of the pulses so as to
reduce the low-frequency parts in the system spectrum and thus its overlap with
the low-frequency portion of the bath spectrum \cite{uhr07}.
DD sequences are inherently \emph{binary}, i.e., their pulsed control parameters
are discretely switched on or off. 
Realistically, the finiteness of pulse durations and spacings sets an upper
limit on the speed and fidelity of DD-assisted quantum gate operations
\cite{vio98,KLi05,uhr07}.

An alternative strategy formulated here in full generality
is \emph{analog unitary control} of multidimensional systems subject to
\emph{any} noise or decoherence. It is effected by a system Hamiltonian whose
time-dependence is variationally tailored to optimally perform a desired gate
operation. The vast additional freedom of non-discrete (smooth) Hamiltonian
parametrization significantly
enhances the efficacy of decoherence control under realistic constraints
compatible with the non-Markov time scales required for such control.
Its formulation meets the long-standing conceptual challenge of
\emph{simultaneously} controlling non-commuting system operators subject to
noise along \emph{orthogonal} axes. This is here achieved by working in an
optimally rotated, different basis at each instant.
The price we pay for such general optimal control is the need for at least
partial knowledge of the bath or noise spectrum, which is \emph{experimentally}
accessible \cite{Sag09} without the need for microscopic models.
The goal is to \emph{minimize its overlap} with the spectrum of the controlled
system, as was already shown for pure dephasing of qubits \cite{gkl08}.
\section{Gate Error}
We assume that the system Hamiltonian $\hat{H}_{\mathrm{S}}(t)$ implements a
desired quantum gate operation at time $t$, and aim at designing it so as to
minimize the decoherence and noise errors. The system-bath interaction
$\hat{H}_{\mathrm{I}}$ then acquires time-dependence in the interaction picture
under the action of $\hat{H}_{\mathrm{S}}(t)$ and the bath Hamiltonian
$\hat{H}_{\mathrm{B}}$.
Assuming factorized initial states of the system and the bath,
$\hat{\varrho}_{\mathrm{tot}}(0)$ $\!=$
$\!\hat{\varrho}(0)\otimes\hat{\varrho}_{\mathrm{B}}$, tracing over the
bath, and further assuming that
$\mathrm{Tr}_{\mathrm{B}}
[\hat{H}_{\mathrm{I}}(t)\hat{\varrho}_{\mathrm{B}}]$
$\!=$ $\!\hat{0}$, yields for the system state $\hat{\varrho}(t)$
the integrated (exact) deviation from the initial state (App.A),
\begin{eqnarray}
  \hat{\varrho}(t)&=&\!\hat{\varrho}(0)-\Delta\hat{\varrho}(t),
  \nonumber\\
  \Delta\hat{\varrho}(t)&=&
  \!\!\int_{0}^{t}\!\!\!\mathrm{d}t_1\!\int_{0}^{t_1}\!\!\!\mathrm{d}t_2
  \mathrm{Tr}_{\mathrm{B}}
  [\hat{H}_{\mathrm{I}}(t_1),
  [\hat{H}_{\mathrm{I}}(t_2),\hat{\varrho}_{\mathrm{tot}}(t_2)]].\quad
\label{NZi}
\end{eqnarray}

In what follows, we assume that up to $t$, the combined system-bath state
changes only weakly compared to $\hat{H}_{\mathrm{I}}$, so that we approximate
in (\ref{NZi})
$\hat{\varrho}_{\mathrm{tot}}(t_2)\approx\hat{\varrho}_{\mathrm{tot}}(0)$
in the integral. This means that the control is assumed effective enough to
allow only \emph{small errors}, consistently with the first order approximation
of the solutions of both the Nakajima-Zwanzig and the time-convolutionless
master equations \cite{JPB,bookBreuer}.

To justify this assumption, we try to reduce the discrepancy between the states
evolved for time $t$ in the presence and absence of the bath by minimizing
$\langle\Delta\hat{\varrho}(t)\rangle$ $\!\equiv$
$\!\langle\Psi|\Delta\hat{\varrho}(t)|\Psi\rangle$ averaged over
\emph{all initial states} $|\Psi\rangle$ that are unknown in general.
For a $d$-level system this averaging is tantamount to taking the expectation
value with respect to the maximum entropy state
$\hat{\varrho}_{\mathrm{S}}$ $\!=$ $\!d^{-1}\hat{I}$.
Assuming that
$\mathrm{Tr}_{\mathrm{S}}[\hat{H}_{\mathrm{I}}(t)\hat{\varrho}_{\mathrm{S}}]$
$\!=$ $\!\hat{0}$, we obtain our measure of decoherence (error) in the form of
(App.A)
\begin{eqnarray}
  \overline{\langle\Delta\hat{\varrho}(t)\rangle}
  &=&\!2\kappa\mathrm{Re}
  \!\int_{0}^{t}\!\mathrm{d}t_1\!\int_{0}^{t_1}\!\mathrm{d}t_2
  \bigl\langle
  \hat{H}_{\mathrm{I}}(t_1)\hat{H}_{\mathrm{I}}(t_2)
  \bigr\rangle_{\mathrm{SB}}
  \nonumber\\
  &=&\kappa
  \Bigl\langle
  \Bigl[\int_{0}^{t}\!\mathrm{d}t_1\hat{H}_{\mathrm{I}}(t_1)\Bigr]^2
  \Bigr\rangle_{\mathrm{SB}},
\label{grate}
\end{eqnarray}
where $\kappa$ $\!=$ $\!1\!-\!(d\!+\!1)^{-1}$,
$\langle\cdot\rangle_{\mathrm{SB}}$ $\!=$
$\!\mathrm{Tr}_{\mathrm{SB}}[(\cdot)\hat{\varrho}_{\mathrm{S}}
\otimes\hat{\varrho}_{\mathrm{B}}]$. Hence,
$\overline{\langle\Delta\hat{\varrho}(t)\rangle}$ is always positive and
proportional to the mean square of the interaction energy as observed in the
interaction picture (by a co-rotating observer).

Since our aim is to suppress $\overline{\langle\Delta\hat{\varrho}(t)\rangle}$
by system manipulations alone, we now separate system and
bath parts by decomposing any interaction Hamiltonian in an orthogonal basis of
system states $|j\rangle$ as
\begin{equation}
\label{HID2}
  \hat{H}_{\mathrm{I}}(t)
  =\sum_{j=1}^{d^2-1}\hat{B}_j(t)\hat{S}_j(t),
\end{equation}
where the Hermitian $\hat{B}_j$ and $\hat{S}_j$ are bath and system operators,
respectively, assumed to obey $\langle\hat{B}_j(t)\rangle_{\mathrm{B}}$ $\!=$
$\!\mathrm{Tr}\hat{S}_j(t)$ $\!=$ $\!0$ and carry no explicit time
dependence. In the interaction picture
\begin{eqnarray}
  \hat{B}_j(t)&=&\mathrm{e}^{\mathrm{i}\hat{H}_{\mathrm{B}}t}\hat{B}_j
  \mathrm{e}^{-\mathrm{i}\hat{H}_{\mathrm{B}}t},
  \nonumber\\
  \hat{S}_j(t)&=&\hat{U}^\dagger(t)\hat{S}_j\hat{U}(t),
  \nonumber\\
\label{Uop}
  \hat{U}(t)&=&\mathrm{T}_+
  \mathrm{e}^{-\mathrm{i}\int_{0}^{t}\!\mathrm{d}t^\prime
  \hat{H}_{\mathrm{S}}(t^\prime)}.
\end{eqnarray}
We shall minimize $\overline{\langle\Delta\hat{\varrho}(t)\rangle}$ for given,
experimentally accessible \cite{Sag09}, bath correlations
\begin{equation}
\label{Phijkconst}
  {\Phi}_{jk}(t)=\bigl\langle\hat{B}_j(t)\hat{B}_k\bigr\rangle_{\mathrm{B}}.
\end{equation}

It is expedient to define the decoherence matrix
\begin{equation}
\label{Rmat1t2}
  \underline{\bm{R}}(t_1,t_2)
  =\underline{\bm{\epsilon}}^T(t_1)\underline{\bm{\Phi}}(t_1-t_2)
  \underline{\bm{\epsilon}}(t_2),
\end{equation}
which obeys
$\underline{\bm{R}}^\dagger(t_1,t_2)$ $\!=$ $\!\underline{\bm{R}}(t_2,t_1)$.
It is the matrix product of the bath
correlation matrix $\underline{\bm{\Phi}}$ formed from the coefficients
${\Phi}_{jk}$ in (\ref{Phijkconst}) and the system-modulation (rotation)
matrix defined as
\begin{eqnarray}
  \hat{S}_j(t)&=&\sum_{k=1}^{d^2-1}{\epsilon}_{jk}(t)\hat{S}_k,
  \nonumber\\
  \quad{\epsilon}_{jk}(t)&=&\frac{1}{2}
  \mathrm{Tr}[\hat{S}_j(t)\hat{S}_k],
\label{Rjk}
\end{eqnarray}
where we have assumed that
$\mathrm{Tr}(\hat{S}_j\hat{S}_k)$ $\!=$ $\!2\delta_{jk}$.
The transformation (\ref{Rjk}) is at the heart of the treatment: it defines the
instantaneous rotating frame where the system and bath are maximally decoupled,
as shown below.

We can now write (\ref{grate}) as (App.B)
\begin{equation}
\label{deltarho}
  \overline{\langle\Delta\hat{\varrho}(t)\rangle}
  =2\frac{\kappa}{d}\;\int_{0}^{t}\!\mathrm{d}t_1\int_{0}^{t}\!\mathrm{d}t_2
  \mathrm{Tr}\underline{\bm{R}}(t_1,t_2).
\end{equation}
Alternatively, we can rewrite (\ref{deltarho}) as
\begin{equation}
\label{som}
  \overline{\langle\Delta\hat{\varrho}(t)\rangle}
  =4t\,\frac{\kappa}{d}\int_{0}^{\infty}\mathrm{d}\omega\,
  \mathrm{Tr}[\underline{\bm{G}}_{}(\omega)\underline{\bm{F}_t}(\omega)],
\end{equation}
i.e., as the spectral overlap of two matrix-valued functions: the
bath coupling spectral matrix $\underline{\bm{G}}_{}(\omega)$ $\!=$
$\!\int_{-\infty}^{\infty}\!\mathrm{d}t\;\mathrm{e}^{\mathrm{i}\omega{t}}\;
\mathrm{Re}\underline{\bm{\Phi}}_{}(t)$, and the system-modulation
spectral matrix at finite time $t$ [cf. (\ref{Rmat1t2})]
$\underline{\bm{F}_t}(\omega)$ $\!=$
$\!\frac{1}{t}\underline{\bm{\epsilon}_t}(\omega)
\underline{\bm{\epsilon}_t}^\dagger(\omega)$,
$\underline{\bm{\epsilon}_t}(\omega)$ $\!=$
$\!\frac{1}{\sqrt{2\pi}}\int_{0}^t\mathrm{d}\tau\,
\mathrm{e}^{\mathrm{i}\omega\tau}\underline{\bm{\epsilon}}(\tau)$.
In (\ref{som}) we have made use of the fact that
$\underline{\bm{\Phi}}(-t)$ $\!=$ $\!\underline{\bm{\Phi}}^\dagger(t)$, so that
it is sufficient to integrate over positive frequencies.

Equation (\ref{som}) constitutes a generalization of the ``universal formula''
\cite{kof01} to arbitrary multidimensional systems and baths.
It provides a major insight: the system and bath spectra (all matrix components)
must be \emph{anticorrelated}, i.e., ${G}_{jk}(\omega)$ minima must coincide
with $({F}_t)_{jk}(\omega)$ maxima and vice versa to minimize (\ref{som}), as
illustrated below.
It should be emphasized that for given $\omega$, both
$\underline{\bm{G}}_{}(\omega)$ and $\underline{\bm{F}_t}(\omega)$ are positive
matrices.
Nevertheless, certain components ${G}_{jk}(\omega)$, $({F}_t)_{jk}(\omega)$ may
be \emph{negative} if $d$ $\!>$ $\!2$ (i.e., not for qubits).
This may allow us to `destructively interfere' their contributions,
i.e., \emph{engineer} ``dark states'' \cite{gor06b} or ``decoherence-free''
subspaces \cite{Zan05}.
These prospects of our general scheme will be explored elsewhere.
\section{Decoherence Minimization}
Our goal is to find a system Hamiltonian $\hat{H}_{\mathrm{S}}({t_1})$,
$0$ $\!\le$ $\!t_1$ $\!\le$ $\!t$, implementing a given unitary gate
$\hat{U}(t)$ at a fixed time $t$ according to (\ref{Uop}).
This requires minimizing (\ref{grate}) or (\ref{deltarho}), 
$\overline{\langle\Delta\hat{\varrho}(t)\rangle}$ $\!\to$ $\!{min}$,
i.e., minimizing the bath-induced state error in the interaction picture under
$\hat{H}_{\mathrm{S}}(t)$.
We may similarly account for the effects of \emph{modulation or control noise},
in addition to bath noise (App.C).

The major difficulty in minimizing (\ref{deltarho}) using
(\ref{Uop})-(\ref{Rjk}) is that (\ref{Uop}) involves time-ordered integration
for arbitrary bath and control axes. To circumvent this
difficulty, we make use of $\hat{U}(t_1)$ instead of
$\hat{H}_{\mathrm{S}}$, and assume a parametrization
$\hat{U}[{f}_l({t_1}),{t_1}]$ in terms of a set of \emph{real parameters}
${{f}}_l({t_1})$, which may be combined to a vector
$\underline{\bm{{f}}}({t_1})$. The number of parameters may vary, since the
parametrization does not have to be complete. The boundary values
$\underline{\bm{{f}}}(0)$ and $\underline{\bm{{f}}}(t)$ should be such
that $\hat{U}(t_1\!=\!0)$ $\!=$ $\!\hat{I}$ and $\hat{U}(t_1\!=\!t)$ is the
desired gate.

If a bath coupling spectrum $\underline{\bm{G}}(\omega)$ vanishes (has cutoff)
at any high frequency, the overlap (\ref{som}) can be presumed arbitrarily
small under sufficiently rapid modulation of the Hamiltonian, such that all
components of $\underline{\bm{F}_t}(\omega)$ are shifted beyond this cutoff,
thus achieving DD \cite{vio98,KLi05,uhr07}. Yet this may require a
\emph{diverging} system energy. Furthermore,
\emph{fidelity generally drops with modulation energy}, as discussed below.
We therefore impose an energy constraint on the modulated system
\begin{equation}
\label{ES}
  E_{\mathrm{S}}=\int_{0}^{t}\!\mathrm{d}{t_1}
  \bigl\langle\hat{H}_{\mathrm{S}}^2({t_1})\bigr\rangle_{\mathrm{S}}={const}.,
\end{equation}
where $\langle\cdot\rangle_{\mathrm{S}}$ $\!=$
$\!\mathrm{Tr}[(\cdot){d}^{-1}\hat{I}]$ [cf. (\ref{grate})]. An alternative
constraint
\begin{equation}
\label{constraint}
  E=\int_{0}^{t}\mathrm{d}{t_1}\;|\dot{\underline{\bm{{f}}}}({t_1})|^2
  ={const}.
\end{equation}
allows a simplified treatment.
In general, $E$ accounts for the fact that the time dependence of a
parametrization cannot be arbitrarily fast and hence bounds the modulated
$\hat{H}_{\mathrm{S}}({t_1})$, thus also limiting $E_{\mathrm{S}}$.

The minimization of (\ref{deltarho}) subject to (\ref{constraint}) is an
extremal problem in terms of $\underline{\bm{{f}}}$.
Denoting by $\underline{\bm{\delta}}$ the total variation with respect to
$\underline{\bm{{f}}}$, the stationary condition can be formulated in terms of a
Lagrange multiplier $\lambda$ as
$\underline{\bm{\delta}}\overline{\langle\Delta\hat{\varrho}(t)\rangle}$ $\!+$
$\!\lambda\underline{\bm{\delta}}E$ $\!=$ $\!0$.
Then, using the parametrization in $\underline{\bm{R}}$ [Eq.~(\ref{Rmat1t2})],
$\nabla\underline{\bm{\epsilon}}$ $\!\equiv$
$\!\{\frac{\partial}{\partial{{f}}_l}\underline{\bm{\epsilon}}(t_1)\}$,
yields the Euler-Lagrange equation
\begin{equation}
\label{elel}
  \ddot{\underline{\bm{{f}}}}(t_1)
  =\lambda\underline{\bm{{g}}}(t_1),
  \quad
  \underline{\bm{{g}}}(t_1)\equiv\int_{0}^{t}\!\mathrm{d}t_2
  \nabla\mathrm{Re}\mathrm{Tr}\underline{\bm{R}}(t_1,t_2),
\end{equation}
where $\lambda$ is related to the constraint (\ref{constraint}) on $E$
(App.D).

We conclude the general treatment by recapitulating on the steps to find the
optimal modulation of $\hat{H}_{\mathrm{S}}({t_1})$:
1) After defining the `cycle time' $t$ and gate operation $\hat{U}(t)$, we
declare a parametrization $\hat{U}[\underline{\bm{{f}}}({t_1}),{t_1}]$ which
induces a parametrization
$\underline{\bm{\epsilon}}[\underline{\bm{{f}}}({t_1}),{t_1}]$ that in turn
yields $\underline{\bm{R}}(t_1,t_2)$ as a functional of $\underline{\bm{{f}}}$
via (\ref{Uop})-(\ref{Rjk}), using our knowledge of (\ref{Phijkconst}).
2) We now solve (\ref{elel}) for a given initial
$\underline{\bm{{f}}}_{\mathrm{init}}(t_1)$ satisfying the boundary conditions,
e.g., such that $\hat{U}[\underline{\bm{{f}}}_{\mathrm{init}}(t_1),{t_1}]$
$\!=$ $\![\hat{U}(t)]^{\frac{t_1}{t}}$, and calculate 
$\overline{\langle\Delta\hat{\varrho}(t)\rangle}$.
3) The optimization is repeated for different values of $\lambda$ and
$E_{\mathrm{S}}$ in (\ref{ES}) is calculated for each
solution.
Among all solutions for which 
$\overline{\langle\Delta\hat{\varrho}(t)\rangle}$ falls below a desired
threshold value, we choose the one corresponding to the lowest $E_{\mathrm{S}}$.
4) The chosen solution $\underline{\bm{{f}}}({t_1})$ is inserted into
$\hat{U}[\underline{\bm{{f}}}({t_1}),{t_1}]$ in (\ref{Uop}), yielding the
instantaneous control parameters
\begin{equation}
\label{instpar}
  \hat{H}_{\mathrm{S}}({t_1})=\sum_j\omega_j({t_1})\hat{S}_j,\quad
  \omega_j({t_1})=\frac{1}{2}\mathrm{Tr}[\hat{S}_j\hat{H}_{\mathrm{S}}({t_1})].
\end{equation}
\section{Application to a qubit}
\begin{figure}[ht]
\includegraphics[width=4.2cm]{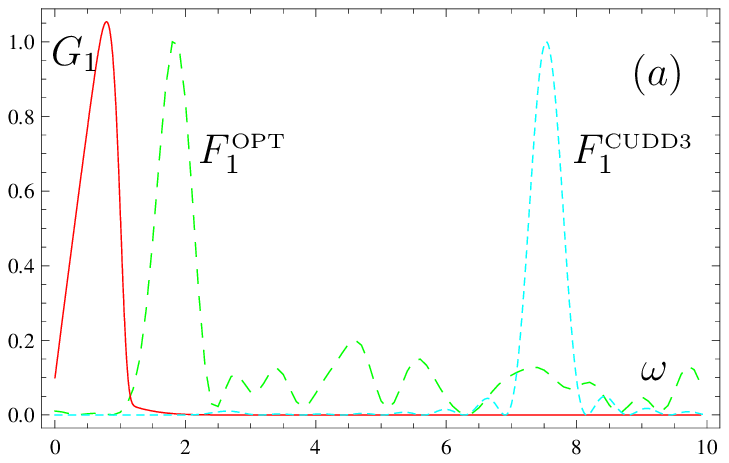}
\includegraphics[width=4.2cm]{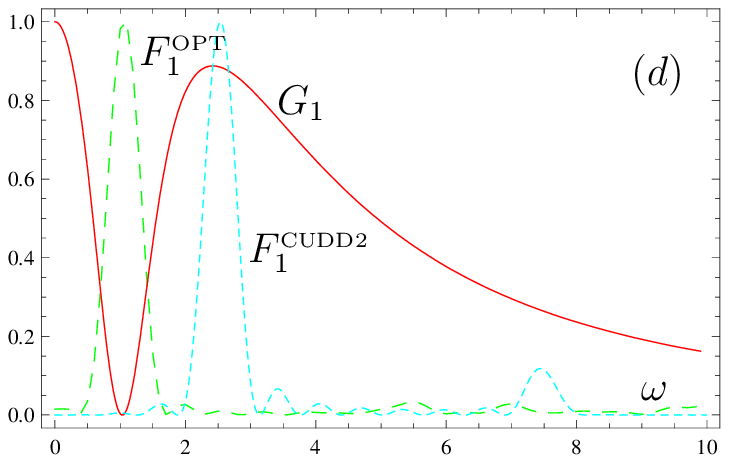}\\
\includegraphics[width=4.2cm]{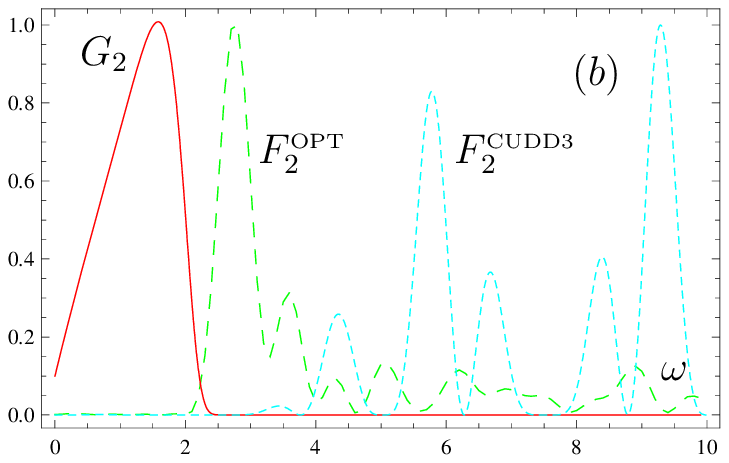}
\includegraphics[width=4.2cm]{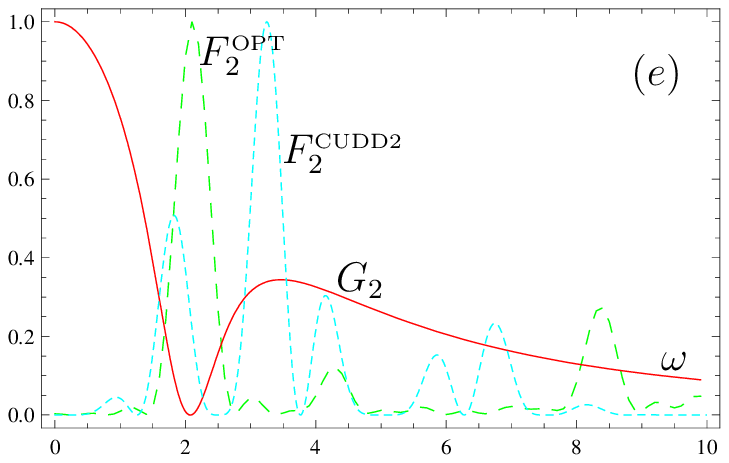}\\
\includegraphics[width=4.2cm]{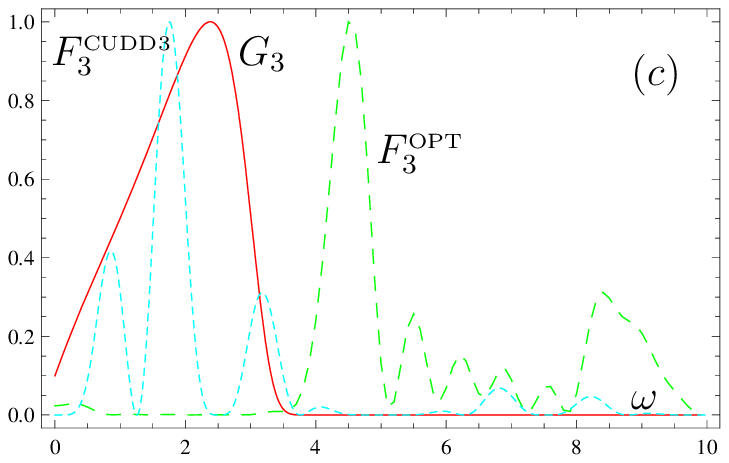}
\includegraphics[width=4.2cm]{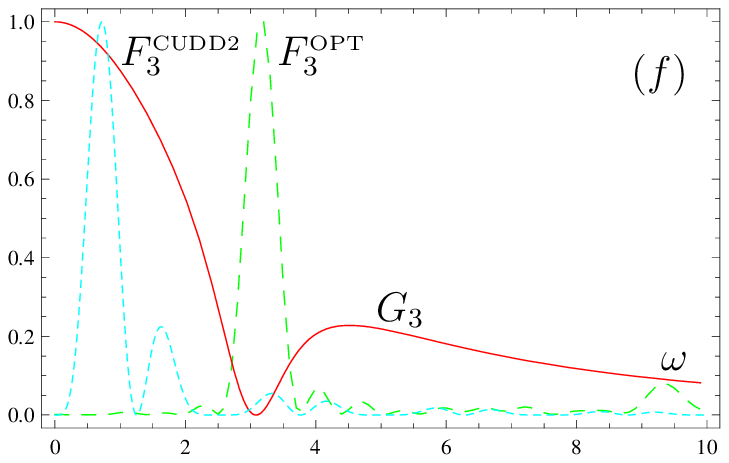}
\caption{\label{F1}
Spectral overlaps between bath spectra $G_i(\omega)$ (solid red),
modulation spectra $F_i^{\scriptscriptstyle{\mathrm{OPT}}}(\omega)$
(dashed green) for an optimized $\pi$-gate at $E_{\mathrm{S}}$ $\!=$ $\!133.2$
[(a),(b),(c)], and an optimized identity ($0$-) gate at
$E_{\mathrm{S}}$ $\!=$ $\!181.1$ [(d),(e),(f)], respectively,
and modulation spectra $F_i^{\scriptscriptstyle{\mathrm{CUDD}}}$ for pulse
sequences (App.F) CUDD3 [(a),(b),(c)] and CUDD2 [(d),(e),(f)]
(dotted blue), with $i=1,2,3$ corresponding to $X$, $Y$, and
$Z$-component, respectively.
Graphs (a),(b),(c) represent an Ohmic bath spectrum with softened cutoff,
whereas graphs (d),(e),(f) represent a Lorentzian spectrum with a dip.
The optimal modulation spectra $F_i^{\scriptscriptstyle{\mathrm{OPT}}}(\omega)$
are always \emph{anticorrelated} with the bath spectra $G_i(\omega)$.
By contrast, Uhrig's pulse sequence spectra
$F_i^{\scriptscriptstyle{\mathrm{CUDD}}}$ are only anticorrelated with $G_i$ for
Ohmic baths (a),(b) but not for the bath spectra (d),(e),(f).
}
\end{figure}

%
To apply the general procedure to a qubit for which
$\hat{S}_j$ $\!=$ $\!\hat{\sigma}_j$
($j$ $\!=$ $\!x,y,z$) in (\ref{instpar}), we resort to the Euler
rotation-angle parametrization,
\begin{displaymath}
  \hat{U}(t)=
  \mathrm{e}^{-\frac{\mathrm{i}}{2}f_3(t)\hat{\sigma}_3}
  \mathrm{e}^{-\frac{\mathrm{i}}{2}f_2(t)\hat{\sigma}_2}
  \mathrm{e}^{-\frac{\mathrm{i}}{2}f_1(t)\hat{\sigma}_3}.
\end{displaymath}
In (\ref{instpar}), $\omega_3(t)$ is now the level splitting, whereas
$\omega_{1(2)}(t)$ are Rabi flipping rates.
We choose two examples of uncorrelated
(i.e., diagonal) baths, namely, an Ohmic bath with \emph{different cutoffs} in
$X$, $Y$, $Z$, and a Lorentzian noise spectrum superposed with a second
Lorentzian such that a spectral `hole' is obtained at different frequencies in
$X$, $Y$, and $Z$. The corresponding bath coupling spectra are shown in
Fig.~\ref{F1}, along with our optimized modulation spectra, which are contrasted
with Uhrig's DD pulse-sequence spectra \cite{uhr07} (App.E,F).

\begin{figure}[ht]
\includegraphics[width=8.6cm]{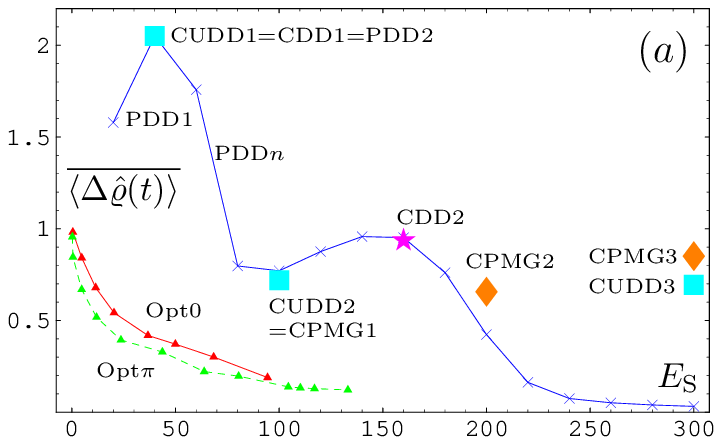}\\
\includegraphics[width=8.6cm]{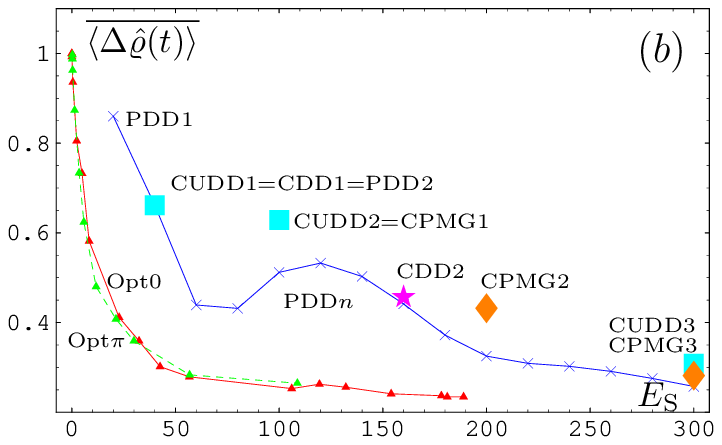}
\caption{\label{F2}
Qubit gate error
$\overline{\langle\Delta\hat{\varrho}(t)\rangle}$ in units
of a reference $\overline{\langle\Delta\hat{\varrho}(t)\rangle}_{\mathrm{ref}}$
which corresponds to a time-independent initial system Hamiltonian (with which
the optimization started), as a function of the constraint $E_{\mathrm{S}}$.
Solid red (dashed green): optimized identity ($\pi$)-gate,
Solid blue: periodic $X$-$Z$-``bang bang'' (2,4,$\ldots$,30 pulses).
Separate points: concatenated Uhrig and related pulse sequences (App.F).
(a) and (b) correspond to bath spectra shown on the left and right in
Fig.~\ref{F1}.
A different scale of $E_{\mathrm{S}}$ is used for pulse sequences whose
$E_{\mathrm{S}}$ is given in units of $E_{\mathrm{S}}^{(1)}/10$, where
$E_{\mathrm{S}}^{(1)}$ $\!=$ $\!\pi^2/(4T)=2.467\cdot10^{3}$ is the single
$\pi$-pulse energy, assuming nearly-ideal square pulses.
}
\end{figure}

The minimized gate error is shown in Fig.~\ref{F2} as a function of the energy
constraint (\ref{ES}) for both baths.
Its comparison with the gate error obtained using various DD pulse sequences
reveals two differences. The first concerns the energy scale:
in rectangular DD pulse sequences, each $\pi$-pulse of duration $T$
contributes an
amount $\pi^2/(4T)$ to (\ref{ES}), which \emph{diverges} for ideal pulses,
$T$ $\!\to$ $\!0$. By contrast, our approach assumes finite, much smaller
$E_{\mathrm{S}}$.
The second difference concerns energy monotonicity:
DD-sequences are designed a priori, regardless of the bath-spectrum, and hence
only significantly reduce the gate error if $E_{\mathrm{S}}$ has risen above
some threshold which is needed to shift all system frequencies beyond the bath
cutoffs \cite{uhr07}. In contrast, our approach starts to reduce the gate error
as soon as $E_{\mathrm{S}}$ $\!>$ $\!0$, since it optimizes the use of the
available energy, by anti-correlating the modulation and bath spectra.

\begin{figure}[ht]
\includegraphics[width=8.6cm]{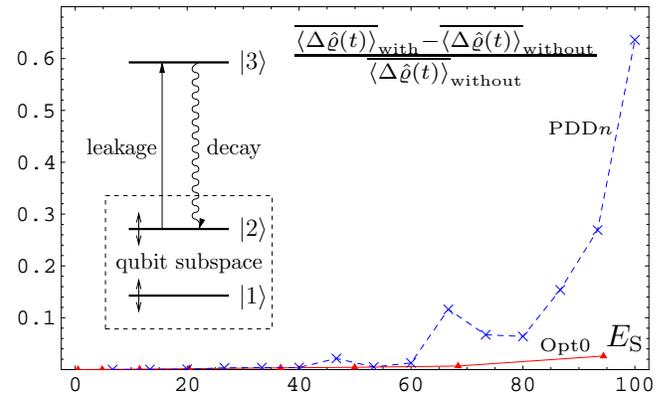}
\caption{\label{F3}
Relative surplus error incurred by the inevitable leakage to the additional
level $|3\rangle$ (inset) caused by control $f(t)$ as a function of
$E_{\mathrm{S}}$, i.e., error with allowance for leakage compared to the error
without (disregarding) leakage (App.G).
Solid red: optimized identity-gate, dashed blue: periodic $X$-$Z$-``bang bang''
(2,4,$\ldots$,30 pulses) as shown in Fig.~\ref{F2}(a) and a bath spectrum as
shown in Fig.~\ref{F1}(a),(b),(c).
A (truncated) $1/\omega$ bath coupling spectrum describes an amplitude coupling
between levels $|2\rangle$ and  $|3\rangle$ (leakage).
As in Fig.~\ref{F2}, a different scale of $E_{\mathrm{S}}$ is used for the pulse
sequences: $E_{\mathrm{S}}$ is here given in units of
$(3/10)E_{\mathrm{S}}^{(1)}$.
}
\end{figure}

We next consider the \emph{gate fidelity limitations as a function of $E_S$}
posed by \emph{leakage} \cite{WKB09} to levels \emph{outside} the relevant
subspace (here a qubit). In a 3-level $\Lambda$-system, any off-resonant control
field acting on the qubit levels $|1\rangle$, $|2\rangle$, causes leakage to the
unwanted level $|3\rangle$ \cite{aga96}. Such leakage and the ensuing incoherent
decay $|3\rangle$ $\!\to$ $\!|1\rangle$ incur gate errors that grow with
$E_{\mathrm{S}}$ (App.G). This behaviour is illustrated in Fig.~\ref{F3},
which reveals that leakage error is the more dramatic, the more energetic
the $\pi$-pulse sequences are. If $\pi$-pulses are experimentally implemented as
$(2m+1)\pi$-pulses, $m{\scriptstyle\gtrapprox}10^3$, it can therefore be
expected that isolated manipulation on a subspace is difficult.
This, together with the qubit-gate optimization, the general expressions for
the gate error (\ref{deltarho}) and (\ref{som}) and its minimization
(\ref{elel}) are the main results of this work.
\section{Conclusions}
A) We have expressed an \emph{arbitrary} gate error for finite-dimensional
quantum systems as the spectral overlap between the driven-system and the bath
spectra.
B) We have derived a non-Markovian Euler-Lagrange equation for the time
dependence of control parameters whose solution maximizes the gate fidelity.
C) This solution leads to \emph{anticorrelation} of the system and bath
spectra. Hence, while DD-based methods rely on shifting the \emph{entire}
spectrum of the system beyond that of the bath,
our optimization takes advantage of gaps or dips of the bath spectra.
D) The treatment of a qubit demonstrates that our approach is significantly more
economic in terms of energy investment than DD-based methods. Such energy saving
may be crucial in terms of fidelity as excessive energies lead to \emph{leakage}
into additional levels \cite{aga96}, or \emph{increase} the control noise
\cite{Sag09}.
\acknowledgments
The support of EC (MIDAS), DIP, ISF and the Humboldt Award (G.K.) is
acknowledged.
\appendix
\section{Derivation of the decoherence (error) expression (\ref{grate})}
The von Neumann equation for the total density operator of the bath and system
combined in the interaction picture,
\begin{equation}
  \frac{\partial}{\partial{t}}\hat{\varrho}_{\mathrm{tot}}(t)
  =-\mathrm{i}[\hat{H}_{\mathrm{I}}(t),\hat{\varrho}_{\mathrm{tot}}(t)],
\label{vne}
\end{equation}
can be written in integrated form
\begin{equation}
  \hat{\varrho}_{\mathrm{tot}}(t)=\hat{\varrho}_{\mathrm{tot}}(0)
  -\mathrm{i}\int_{0}^{t}\!\mathrm{d}t_1
  [\hat{H}_{\mathrm{I}}(t_1),\hat{\varrho}_{\mathrm{tot}}(t_1)].
\label{vnei}
\end{equation}
Substituting (\ref{vnei}) back into (\ref{vne}) gives
\begin{eqnarray}
  \frac{\partial}{\partial{t}}\hat{\varrho}_{\mathrm{tot}}(t)
  &=&-\mathrm{i}[\hat{H}_{\mathrm{I}}(t),\hat{\varrho}_{\mathrm{tot}}(0)]
  \nonumber\\
  &&-\!\int_{0}^{t}\!\mathrm{d}t_1
  [\hat{H}_{\mathrm{I}}(t),
  [\hat{H}_{\mathrm{I}}(t_1),\hat{\varrho}_{\mathrm{tot}}(t_1)]],\quad
\end{eqnarray}
and after tracing over the bath,
\begin{equation}
\label{ME}
  \frac{\partial}{\partial{t}}\hat{\varrho}(t)
  =-\!\int_{0}^{t}\!\mathrm{d}t_1\mathrm{Tr}_{\mathrm{B}}
  [\hat{H}_{\mathrm{I}}(t),
  [\hat{H}_{\mathrm{I}}(t_1),\hat{\varrho}_{\mathrm{tot}}(t_1)]].
\end{equation}
Although we do not make use of the differential equation for the system state
$\hat{\varrho}(t)$, it may be useful to mention that it can be obtained from
(\ref{ME}) by neglecting the bath correlations, i.e., setting
$\hat{\varrho}_{\mathrm{tot}}(t_1)$ $\!\approx$
$\!\hat{\varrho}_{\mathrm{B}}\otimes\hat{\varrho}(t_1)$, which yields the
second-order Nakajima-Zwanzig equation \cite{bookBreuer}.
Replacing $\hat{\varrho}_{\mathrm{tot}}(t_1)$
$\!\approx$ $\!\hat{\varrho}_{\mathrm{B}}\otimes\hat{\varrho}(t)$ instead
yields the second-order time-convolutionless equation.
For the averaging, we make use of
\begin{equation}
\label{dankert}
  \overline{\langle\Psi|\hat{A}|\Psi\rangle\langle\Psi|\hat{B}|\Psi\rangle}
  =\frac{\mathrm{Tr}\hat{A}\hat{B}+\mathrm{Tr}\hat{A}\mathrm{Tr}
  \hat{B}}{d(d+1)}
\end{equation}
\cite{dan05} to write the covariance of two operators $\hat{A}$ and $\hat{B}$ as
\begin{eqnarray}
  Cov(\hat{A},\hat{B})&\equiv&
  \overline{
  \langle\Psi|\hat{A}\hat{B}|\Psi\rangle
  -\langle\Psi|\hat{A}|\Psi\rangle\langle\Psi|\hat{B}|\Psi\rangle
  }
  \nonumber\\
  &=&\kappa\bigl(\langle\hat{A}\hat{B}\rangle
  -\langle\hat{A}\rangle\langle\hat{B}\rangle\bigr).
\label{cov2}
\end{eqnarray}
Expressing now the double commutator in (\ref{NZi}) as
$[\hat{H}_{\mathrm{I}}(t_1),[\hat{H}_{\mathrm{I}}(t_2),
\hat{\varrho}_{\mathrm{tot}}]]$ $\!=$
$\!([\hat{H}_{\mathrm{I}}(t_1),
\hat{H}_{\mathrm{I}}(t_2)\hat{\varrho}_{\mathrm{tot}}]$ $\!+$ $\!h.a.)$,
and applying (\ref{cov2}), we obtain (\ref{grate}).
\section{Derivation of the spectral overlap error (\ref{som})}
The differential equation (\ref{ME}) for the system state can be written as
[see comments following (\ref{ME})]
\begin{equation}
\label{MEop}
  \frac{\partial}{\partial{t}}\hat{\varrho}(t)
  \!=\!-\sum_{j,k=1}^{d^2-1}\!\int_{0}^{t}\!\mathrm{d}t_1\!
  \bigl\{\!{\Phi}_{jk}(t\!-\!t_1)
  [\hat{S}_j(t),\hat{S}_k(t_1)\hat{\varrho}(t)]\!+\!h.a.\!\bigr\}\!,
\end{equation}
while (\ref{NZi}) reads
\begin{eqnarray}
  \Delta\hat{\varrho}(t)&=&
  \sum_{j,k=1}^{d^2-1}
  \int_{0}^{t}\!\!\!\mathrm{d}t_1\!\int_{0}^{t_1}\!\!\!\mathrm{d}t_2
  \nonumber\\
  &&\times\,
  \{{\Phi}_{jk}(t_1\!-\!t_2)[\hat{S}_j(t_1),\hat{S}_k(t_2)\hat{\varrho}(0)]
  +h.a.\}\quad\quad
\label{dr}
  \\
  &=&
  \sum_{j,k=1}^{d^2-1}\int_{0}^{t}\!\mathrm{d}t_1\int_{0}^{t_1}\!\mathrm{d}t_2
  \nonumber\\
  &&\times\,
  \left\{{R}_{jk}(t_1,t_2)
  [\hat{S}_j,\hat{S}_k\hat{\varrho}(0)]+h.a.\right\}.
  \nonumber
\end{eqnarray}
We can define a decoherence operator
\begin{equation}
\label{Ropt1t2}
  \hat{R}(t_1,t_2)
  =\sum_{j,k=1}^{d^2-1}\hat{S}_j(t_1){\Phi}_{jk}(t_1-t_2)\hat{S}_k(t_2),
\end{equation}
which obeys $\hat{R}^\dagger(t_1,t_2)$ $\!=$ $\!\hat{R}(t_2,t_1)$.
Assuming finite $d$, (\ref{grate}) then becomes
\begin{eqnarray}
  \overline{\langle\Delta\hat{\varrho}(t)\rangle}
  &=&2\frac{\kappa}{d}\mathrm{Re}
  \int_{0}^{t}\mathrm{d}t_1\int_{0}^{t_1}\mathrm{d}t_2
  \mathrm{Tr}\hat{R}(t_1,t_2)
  \nonumber\\
  &=&\frac{\kappa}{d}\int_{0}^{t}\mathrm{d}t_1\int_{0}^{t}\mathrm{d}t_2
  \mathrm{Tr}\hat{R}(t_1,t_2).
\label{overl}
\end{eqnarray}
Alternatively, by defining the spectral counterparts of the ingredients of
(\ref{Ropt1t2}):
\begin{eqnarray}
\label{bcs}
  {G}_{jk}(\omega)
  &=&\int_{-\infty}^{\infty}\!\mathrm{d}t\;\mathrm{e}^{\mathrm{i}\omega{t}}\;
  \mathrm{Re}{\Phi}_{jk}(t),
  \\
\label{Sjomega}
  \hat{S}_j(\omega)&=&\frac{1}{\sqrt{2\pi}}\int_{0}^t\mathrm{d}\tau\,
  \mathrm{e}^{\mathrm{i}\omega\tau}\hat{S}_j(\tau),
  \\
  {F}_{kj}(\omega)&=&\frac{1}{2t}
  \mathrm{Tr}[\hat{S}_k(\omega)\hat{S}_j^\dagger(\omega)],
\label{Fjkomega}
\end{eqnarray}
Equation (\ref{overl}) can be written as the following spectral overlap
\begin{eqnarray}
  \overline{\langle\Delta\hat{\varrho}(t)\rangle}
  &=&2\frac{\kappa}{d}\int_{0}^{\infty}\mathrm{d}\omega\,
  \sum_{j,k=1}^{d^2-1}\mathrm{Tr}
  [\hat{S}_j^\dagger(\omega){G}_{jk}(\omega)\hat{S}_k(\omega)]\quad
  \nonumber\\
  &=&4t\,\frac{\kappa}{d}\int_{0}^{\infty}\mathrm{d}\omega\,
  \sum_{j,k=1}^{d^2-1}{G}_{jk}(\omega){F}_{kj}(\omega).
\label{soo}
\end{eqnarray}
\section{Modulation Errors}
Since, in practice, a modulation can be realized only with finite accuracy, it
is important to consider the effect of modulation errors. To do so, we add to
$\hat{H}_{\mathrm{S}}(t)$ a small random Hamiltonian $\hat{H}_{\mathrm{N}}(t)$
which acts on the system variables and repeat the previous analysis without
$\hat{H}_{\mathrm{N}}(t)$ in the interaction picture. In addition,
we now perform an ensemble average (also denoted with an overbar) over different
realizations of $\hat{H}_{\mathrm{N}}(t)$. Neglecting systematic errors,
$\overline{\hat{H}_{\mathrm{N}}}(t)$ $\!=$ $\!0$, we can in analogy to
(\ref{Phijkconst}) define a correlation matrix
$\underline{\bm{\Phi}}^{\mathrm{N}}(t_1,t_2)$ with elements
\begin{eqnarray}
  {\Phi}^{\mathrm{N}}_{jk}(t_1,t_2)&=&\overline{h_j(t_1)h_k(t_2)},
  \\
  h_j(t)&=&\frac{1}{2}\mathrm{Tr}[\hat{H}_{\mathrm{N}}(t)\hat{S}_j],
\end{eqnarray}
which gives rise to a noise contribution
\begin{equation}
  \underline{\bm{R}}^{\mathrm{N}}(t_1,t_2)
  =\underline{\bm{\epsilon}}^T(t_1)\underline{\bm{\Phi}}^{\mathrm{N}}(t_1,t_2)
  \underline{\bm{\epsilon}}(t_2),
\end{equation}
that must be added to (\ref{Rmat1t2}) with $\underline{\bm{\epsilon}}(t)$
defined as before. Assuming 
${\Phi}^{\mathrm{N}}_{jk}(t_2,t_1)$ $\!=$ $\!{\Phi}^{\mathrm{N}}_{kj}(t_1,t_2)$,
we have $\underline{\bm{R}}^{\mathrm{N}\,\dagger}(t_1,t_2)$ $\!=$
$\!\underline{\bm{R}}^{\mathrm{N}}(t_2,t_1)$, and
(\ref{deltarho}) now holds for
$\overline{\overline{\langle\Delta\hat{\varrho}(t)\rangle}}$: the double
overbar means that 
$\langle\Delta\hat{\varrho}(t)\rangle$ is averaged over both the initial states
and the ensemble. This analysis accounts for modulation errors if we use a
\emph{modified} correlation function containing both system-noise and bath
contributions and refer to the \emph{ensemble} only.
\section{Euler-Lagrange Variational Analysis}
The minimization of (\ref{deltarho}) subject to (\ref{ES}) constitutes the
\emph{original (unsimplified) extremal problem in terms of
$\underline{\bm{{f}}}$}. The stationary condition corresponding to (\ref{ES}),
\begin{equation}
  \underline{\bm{\delta}}\overline{\langle\Delta\hat{\varrho}(t)\rangle}
  +\lambda\underline{\bm{\delta}}E_{\mathrm{S}}=0,
\end{equation}
with variations fixed at the boundaries,
$\underline{\bm{\delta}}\underline{\bm{{f}}}({t_1})|_{{t_1}=0,t}=0$,
yields an Euler-Lagrange-equation
\begin{equation}
\label{ELG}
  \mathrm{Re}\mathrm{Tr}\left[\ddot{\hat{U}}(t_1)\nabla\hat{U}^\dagger(t_1)
  -\lambda\int_{0}^{t}\!\mathrm{d}t_2\nabla\hat{R}(t_1,t_2)\right]=0.
\end{equation}
Here $\nabla_l$ $\!=$ $\!\partial/\partial{{f}}_l(t_1)$ and the double dots
denote a second derivative with regard to $t_1$. In order to obtain (\ref{ELG}),
we have applied in (4) the relation
\begin{equation}
\label{subs}
  \hat{H}_{\mathrm{S}}({t_1})
  =\mathrm{i}\dot{\hat{U}}({t_1})\hat{U}^\dagger({t_1}).
\end{equation}

The Lagrange multiplier in (\ref{elel}) can be shown to obey
\begin{equation}
\label{ele}
  \lambda=\frac{\sqrt{b^2+a(E-c)}-b}{a},
\end{equation}
where
\begin{eqnarray}
  a&=&\int_{0}^{t}\mathrm{d}t_1
  \Bigl|\int_{0}^{t_1}\!\mathrm{d}t_2\,\underline{\bm{{g}}}(t_2)\Bigr|^2,
  \\
  b&=&\int_{0}^{t}\mathrm{d}t_1\int_{0}^{t_1}\!\mathrm{d}t_2\,
  \dot{\underline{\bm{{f}}}}(0)\cdot\underline{\bm{{g}}}(t_2),
  \\
  c&=&t|\dot{\underline{\bm{{f}}}}(0)|^2.
\end{eqnarray}
Note that for $\dot{\underline{\bm{{f}}}}(0)$ $\!=$ $\!0$ we have
$b$ $\!=$ $\!c$ $\!=$ $\!0$ and (\ref{ele}) reduces to $\lambda$ $\!=$
$\!\sqrt{E/a}$.
\section{Bloch Equation Analysis}
\begin{figure}[ht]
\includegraphics[width=8.6cm]{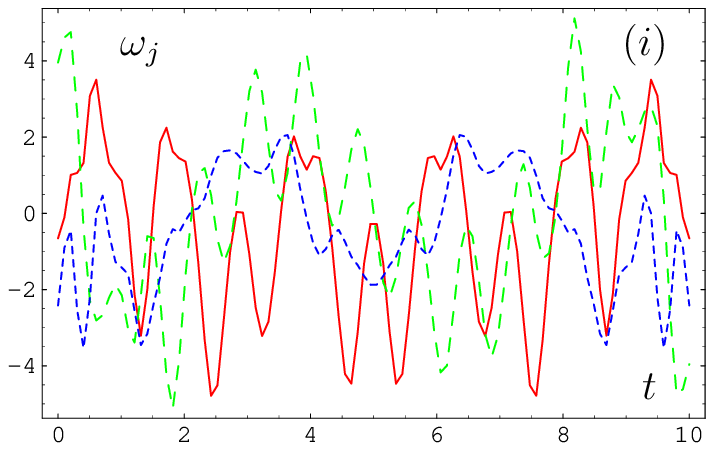}\\
\includegraphics[width=8.6cm]{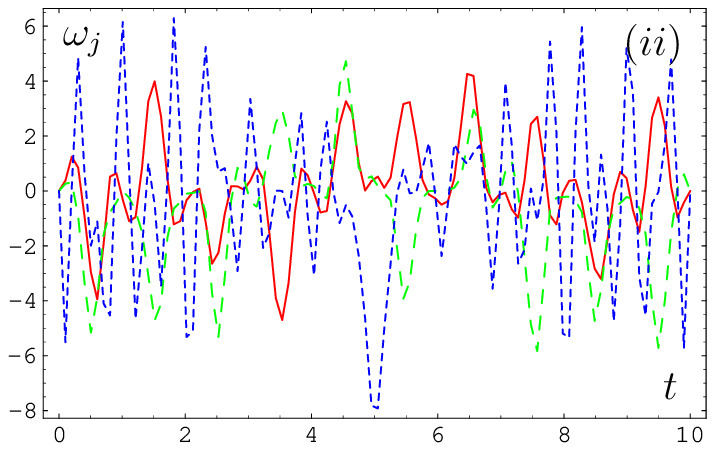}
\caption{\label{F1S}
Optimized time dependence of the control parameters of the system Hamiltonian,
the solid red, dashed green, and dotted blue line show $\omega_{1,2,3}$,
respectively: (i) parameters referring to the graphs (a),(b),(c) in Fig.1;
(ii) parameters referring to graphs (d),(e),(f) in Fig.1.
}
\end{figure}

The state evolution of a qubit can be formulated in terms of the Bloch vector
$\underline{\bm{r}}$ with components
$r_j$ $\!=$ $\!\mathrm{Tr}[\hat{\sigma}_j\hat{\varrho}(t)]$,
$j$ $\!=$ $\!1,2,3$, as the equation of a ``top'' forced by time-dependent
torque
\begin{equation}
\label{BE3}
  \dot{\underline{\bm{r}}}
  =\underline{\bm{L}}_{-}\!\cdot\!\underline{\bm{r}}
  +\underline{\bm{L}}_{+}\!\cdot\!(\underline{\bm{r}}-\underline{\bm{r}}_0).
\end{equation}
Here the matrix function
\begin{equation}
\label{BE3L2}
  \underline{\bm{L}}=4\mathrm{Re}\int_{0}^{t}\!\mathrm{d}t_1
  \{\underline{\bm{R}}^{T}(t,t_1)
  -[\mathrm{Tr}\underline{\bm{R}}(t,t_1)]\bm{I}\}
\end{equation}
has been decomposed into its (anti)symmetric parts
$\underline{\bm{L}}_{\pm}$ $\!=$
$\!(\underline{\bm{L}}$ $\!\pm$ $\!\underline{\bm{L}}^{T})/2$, while 
\begin{eqnarray}
  \underline{\bm{r}}_0
  &=&-\underline{\bm{L}}_{+}^{-1}\!\cdot\!\underline{\bm{b}},
  \\
  {b}_j&=&4\mathrm{Im}\int_{0}^{t}\!\mathrm{d}t_1
  \mathrm{Tr}[\underline{\bm{\sigma}}_j\underline{\bm{R}}(t,t_1)],
\end{eqnarray}
is the quasi-steady state under the chosen time-dependent control. The term
$\underline{\bm{L}}_{+}\!\cdot\!(\underline{\bm{r}}-\underline{\bm{r}}_0)$
accounts for the dynamically-modified relaxation of
$\langle\hat{\sigma}_j\rangle$ at non-Markov time-dependent rates
that are the
eigenvalues of $\underline{\bm{L}}_{+}(t)$, reverting to the standard (Markov)
rates $1/T_j$ in the limit of slow control.
The term $\underline{\bm{L}}_{-}\!\cdot\!\underline{\bm{r}}$ $\!=$
$\!\underline{\bm{\Delta\omega}}\times\underline{\bm{r}}$,
reflects a bath-induced energy shift
\begin{equation}
  {\Delta\omega}_j=2\mathrm{Re}\int_{0}^{t}\!\mathrm{d}t_1
  \mathrm{Tr}[\underline{\bm{\sigma}}_j\underline{\bm{R}}(t,t_1)],
\end{equation}
since it represents a unitary evolution observed in the instantaneous
interaction picture.
The elements of the SO(\ref{HID2}) generator matrices
$\underline{\bm{\sigma}}_j$ can be calculated from
\begin{equation}
  2(\underline{\bm{\sigma}}_j)_{ik}
  =\frac{\mathrm{Tr}([\hat{\sigma}_i,\hat{\sigma}_j]\hat{\sigma}_k)}
  {2\mathrm{i}}.
\end{equation}
The optimized instantaneous control parameters $\omega_j(t)$ are obtained upon
minimizing the departure of $\underline{\bm{r}}$ from its initial value and
following the procedure in the main text leading to (\ref{instpar}). The results
are illustrated in Fig.~\ref{F1S} (see also Fig.~\ref{F1} main text).
\section{Comparison with Uhrig's DD-sequence}
In Figs.~\ref{F1} and \ref{F2} of the main text we compare our results with the
following DD sequences:
\begin{itemize}
\item[]{a)}
Concatenated DD (CDD) \cite{KLi05} defined by
\begin{equation}
  p_{n+1}=p_nXp_nZp_nXp_nZ
\end{equation}
with $p_0$ $\!=$ $\!f_\tau$ denoting free evolution over time $\tau$, where
$p^{\mathrm{CDD}}_1$ $\!=$ $\!(fXfZ)^2$ recovers periodic DD (PDD).
\item[]{b)}
Uhrig-DD (UDD) \cite{uhr07} defined (for $n$ pulses in $Z$)
by
\begin{equation}
  p^{\mathrm{UDD}}_n=f_{t-\tau_n}{Z}f_{\tau_n-\tau_{n-1}}
  {Z}\cdots{Z}f_{\tau_2-\tau_1}{Z}f_{\tau_1}
\end{equation}
with
\begin{equation}
  \tau_{i}=t\sin^2[\pi{j}/(2(n+1))],
\end{equation}
where
\begin{equation}
  p^{\mathrm{UDD}}_1=fZf
\end{equation}
recovers the spin echo (SE) and 
\begin{equation}
  p^{\mathrm{UDD}}_2=(fZf)^2
\end{equation}
the CPMG-sequence.
\item[]{c)}
Combined CDD and UDD in concatenated UDD (CUDD) \cite{uhr07} defined by
concatenating according to
\begin{equation}
  p_{n+1}=p_nXp_nX
\end{equation}
an $m$-pulse UDD sequence
\begin{equation}
  p_0=p^{\mathrm{UDD}}_m
\end{equation}
for $m$ times.
\end{itemize}
The named basic sequences can be iterated, i.e., repeatedly applied.
\section{Leakage from a Subspace}
We can adapt our formalism to the situation where the 
$d$-dimensional state space (to which the relevant quantum information is to be
confined) is a \emph{subspace} of a $N$-dimensional system state
space [12]. To do so, the averaging of the initial states $|\Psi\rangle$ is
performed on the subspace, for which
$|\Psi\rangle$ $\!=$ $\!\hat{P}|\Psi\rangle$, where 
$\hat{P}$ $\!=$ $\!\sum_{n=1}^{d}|\varphi_n\rangle\langle\varphi_n|$
is the associated projector.
Applying (\ref{dankert}) to (\ref{dr}) and defining a matrix
$\underline{\bm{\Gamma}}$ $\!=$ $\!\underline{\bm{\Gamma}}^\dagger$ with
elements
\begin{equation}
  {\Gamma}_{ik}
  =\frac{\mathrm{Tr}(\hat{S}_i\hat{P}\hat{S}_k)}{d}
  -\frac{
  \mathrm{Tr}(\hat{S}_i\hat{P}\hat{S}_k\hat{P})
  +\mathrm{Tr}(\hat{S}_i\hat{P})\mathrm{Tr}(\hat{S}_k\hat{P})
  }{d(d+1)},
\end{equation}
generalizes (\ref{deltarho}) to
\begin{equation}
\label{deltarhogen}
  \overline{\langle\Delta\hat{\varrho}(t)\rangle}
  =\int_{0}^{t}\!\mathrm{d}t_1\int_{0}^{t}\!\mathrm{d}t_2
  \mathrm{Tr}\left[\underline{\bm{R}}(t_1,t_2)\underline{\bm{\Gamma}}\right],
\end{equation}
which recovers (\ref{deltarho}) for $N$ $\!=$ $\!d$, where ${\Gamma}_{ik}$ $\!=$
$\!\frac{2}{d+1}\delta_{ik}$ $\!=$ $\!2\frac{\kappa}{d}\delta_{ik}$.
Equivalently,
\begin{eqnarray}
  \overline{\langle\Delta\hat{\varrho}(t)\rangle}
  &=&2\int_{0}^{\infty}\mathrm{d}\omega\,
  \mathrm{Tr}\bigl[
  \underline{\bm{\epsilon}_t}^\dagger(\omega)
  \underline{\bm{G}}_{\mathrm{re}}(\omega)
  \underline{\bm{\epsilon}_t}(\omega)
  \mathrm{Re}\underline{\bm{\Gamma}}
  \nonumber\\
  &&\quad\quad\quad-\underline{\bm{\epsilon}_t}^\dagger(\omega)
  \underline{\bm{G}}_{\mathrm{im}}(\omega)
  \underline{\bm{\epsilon}_t}(\omega)
  \mathrm{Im}\underline{\bm{\Gamma}}\bigr]
  \nonumber\\
  &=&t\,\int_{-\infty}^{\infty}\mathrm{d}\omega\,
  \mathrm{Tr}[\underline{\bm{G}}_{\mathrm{tot}}(\omega)
  \underline{\bm{F}_t^{\scriptscriptstyle{\Gamma}}}(\omega)],
\label{somtot}
\end{eqnarray}
which replaces (9).
While $\underline{\bm{G}}_{\mathrm{re}}(\omega)$ $\!=$
$\!\int_{-\infty}^{\infty}\!\mathrm{d}t\;\mathrm{e}^{\mathrm{i}\omega{t}}\;
\mathrm{Re}\underline{\bm{\Phi}}_{}(t)$
is identical to $\underline{\bm{G}}_{}(\omega)$ in (9),
here we also need
$\underline{\bm{G}}_{\mathrm{im}}(\omega)$ $\!=$
$\!\int_{-\infty}^{\infty}\!\mathrm{d}t\;\mathrm{e}^{\mathrm{i}\omega{t}}\;
\mathrm{Im}\underline{\bm{\Phi}}_{}(t)$,
or the combined
$\underline{\bm{G}}_{\mathrm{tot}}(\omega)$ $\!=$
$\!\int_{-\infty}^{\infty}\!\mathrm{d}t\;\mathrm{e}^{\mathrm{i}\omega{t}}\;
\underline{\bm{\Phi}}_{}(t)$,
whereas in (\ref{somtot})
$\underline{\bm{F}_t^{\scriptscriptstyle{\Gamma}}}(\omega)$ $\!=$
$\!\frac{1}{t}\underline{\bm{\epsilon}_t}(\omega)\underline{\bm{\Gamma}}
\underline{\bm{\epsilon}_t}^\dagger(\omega)$ replaces
$\underline{\bm{F}_t}(\omega)$ in (9).

(\ref{deltarhogen}) encompasses both the \emph{internal} decoherence effects
within the system-subspace associated with $\hat{P}$ and \emph{leakage} effects
related to a population $\overline{\langle\hat{Q}\rangle}$
$\!=$ $\!\overline{\mathrm{Tr}\bigl[\hat{\varrho}(t)\hat{Q}\bigr]}$ of the
orthogonal complement $\hat{Q}$ $\!=$ $\!\hat{I}$ $\!-$ $\!\hat{P}$, averaged
over all initial states on $\hat{P}$,
\begin{eqnarray}
  \overline{\mathrm{Tr}\bigl[\hat{\varrho}(t)\hat{Q}\bigr]}
  &=&\mathrm{Tr}\bigl[
  \overline{\Delta\hat{\varrho}(t)}\hat{P}\bigr]
  \\
  &=&\int_{0}^{t}\!\mathrm{d}t_1\int_{0}^{t}\!\mathrm{d}t_2
  \mathrm{Tr}\left[\underline{\bm{R}}(t_1,t_2)
  \underline{\bm{\Gamma}_{\mathrm{L}}}\right],
\end{eqnarray}
$\underline{\bm{\Gamma}_{\mathrm{L}}}$ $\!=$
$\!\underline{\bm{\Gamma}_{\mathrm{L}}}^\dagger$ being a matrix with elements
\begin{equation}
  ({\Gamma}_{\mathrm{L}})_{ik}
  =\frac{\mathrm{Tr}(\hat{S}_i\hat{P}\hat{S}_k\hat{Q})}{d}.
\end{equation}
If leakage is disregarded in the procedure minimizing
$\overline{\langle\Delta\hat{\varrho}(t)\rangle}$, it is likely that a
stronger system modulation increases the population of $\hat{Q}$, giving rise
to a significant surplus error. This is illustrated in Fig.3, where optimal and
PDD-modulations originally designed within a two-level model [as
shown in Fig.2(a)] are reconsidered for a two-level subspace of a three-level
system. This is done by replacing the Pauli matrices $\hat{\sigma}_i$ with the
corresponding Gell-Mann matrices $\hat{\gamma}_i$, multiplying
$\hat{U}(t)$ with $\mathrm{e}^{-\mathrm{i}t{f}\hat{\gamma}_8}$ to separate
the levels, and adding to $\hat{H}_{\mathrm{I}}$ a leakage term
$\hat{\gamma}_6\hat{B}_{\mathrm{L}}$. The latter gives rise to an additional
bath correlation function $\Phi_{\mathrm{L}}$, assuming here that it can be
described by a $1/\omega$-bath coupling spectrum.
The total system space is hence spanned by the energy states $|1\rangle$,
$|2\rangle$, and $|3\rangle$, the projector onto the relevant subspace is
$\hat{P}$ $\!=$ $\!\sum_{n=1}^{2}|{n}\rangle\langle{n}|$, whereas 
$\hat{Q}$ $\!=$ $\!|3\rangle\langle3|$, and the states $|\Psi\rangle$ used for
averaging are arbitrary superpositions of $|1\rangle$ and $|2\rangle$.
The time-independent ${f}$ is a parameter that controls the coupling to the
``leakage bath''. It reflects the fact that the energy of the leakage level
$|3\rangle$ induces a free evolution, which is shifted to high
frequencies for sufficiently large ${f}$, when $|3\rangle$ is strongly
energy-detuned from the other two levels, thus providing a ``natural'' dynamic
decoupling of our $1/\omega$-coupling spectrum, and hence the vanishing of the
surplus error induced by leakage, justifying the two-level system approximation.


\end{document}